\documentclass[12pt]{iopart}

\usepackage{graphicx}
\newcommand{\simg}{\stackrel{>}{_\sim}}
\newcommand{\siml}{\stackrel{<}{_\sim}}
\begin{document}

\title[Electronic State of Two Dimensional Triangular Lattice d-p Model for Na$_{x}$CoO$_{2}$]{Electronic State of Na$_{x}$CoO$_{2}$ Based on the Two Dimensional Triangular Lattice d-p Model}

\author{Y. Yamakawa$^{1}$ and Y. \={O}no$^{1,2}$}

\address{$^{1}$ Department of Physics, Niigata University, Ikarashi, Niigata 950-2181, Japan}
\address{$^{2}$ Center for Transdisciplinary Research, Niigata University, Ikarashi, Niigata 950-2181, Japan}
\ead{f05a012j@mail.cc.niigata-u.ac.jp}
\begin{abstract}
The electronic state in a CoO$_2$ plane of the layered cobalt oxides Na$_{x}$CoO$_2$ is investigated by using the 11 band $d$-$p$ model on a two-dimensional triangular lattice, where the tight-binding parameters are determined so as to fit the LDA band structure. Effects of the Coulomb interaction at a Co site: the intra- and inter-orbital direct terms $U$ and $U'$, the exchange coupling $J$ and the pair-transfer $J'$, are treated within the Hartree-Fock approximation. We also consider the effect of the Na order at $x=0.5$, where Na ions form one dimensional chains, by taking into account of an effective one-dimensional potential $\Delta \varepsilon_{d}$ on the CoO$_2$ plane. It is found that the Na order enhances the Fermi surface nesting resulting in the antiferromagnetism (AFM) which is suppressed due to the frustration effect in the case without the Na order. When $U$ and $\Delta \varepsilon_{d}$ are varied, we observe three types of the AFM: (1) the metallic AFM with large density of states $N_F$ at the Fermi level for small values of $U$ and $\Delta \varepsilon_{d}$, (2) the semimetallic AFM with tiny $N_F$ for large $U$ with small $\Delta \varepsilon_{d}$ and (3) the insulating AFM with a finite energy gap for large values of $U$ and $\Delta \varepsilon_{d}$. 
\end{abstract}

\vspace{2pc}
\maketitle

\clearpage

\section{Introduction}
Recently, magnetism and superconductivity in the two dimensional triangular lattice with geometrical frustration have been received much attention. An interesting class of materials is the layered cobalt oxide Na$_x$CoO$_2$, in which the CoO$_2$ plane consists of a triangular lattice of Co atoms. The electron filling in the CoO$_2$ plane is controlled by changing the Na content $x$: the hole concentration of Co $t_{2g}$ bands is given by $n_{\rm hole}=1-x$. Na$_x$CoO$_2$ with $x \simg 0.75$ exhibits a weak magnetic order below 20K \cite{Motohashi}, where the ferromagnetic ordered CoO$_2$ layers couple antiferromagnetically with each other \cite{Bayrakci}. An anomalous metallic state is observed for $0.6\siml x \siml 0.75$: the magnetic susceptibility is Curie-Weiss like \cite{Foo}, $T$-linear coefficient of the specific heat is large and increases with $x$ \cite{Yokoi}, the thermopower is unusually large \cite{Terasaki}. When H$_2$O is intercalated between the CoO$_2$ layers, Takada {\it et al.} \cite{Takada} have discovered the superconductivity in Na$_{x}$CoO$_2\cdot y$H$_2$O with $T_c \sim 5$ K for $x \approx 0.35$ and $y \approx 1.3$. 

Furthermore, Na$_{0.5}$CoO$_2$ exhibits remarkable two transitions at $T_{c1}\sim 87$ K and  $T_{c2}\sim 53$ K \cite{Foo,Yokoi}: the in-plane antiferromagnetic order is realized below $T_{c1}$ and the system becomes insulator below $T_{c2}$. 
From the NMR and the Neutron measurements, Yokoi {\it et al.} \cite{Yokoi} have proposed the magnetic structure of Na$_{0.5}$CoO$_2$, where chains of Co$^{3.5+\delta}$ with larger staggered moment and Co$^{3.5-\delta}$ with smaller moment exist alternatively within the CoO$_2$ plane.  The charge ordering of the Co sites into chains of Co$^{3.5+\delta}$ and Co$^{3.5-\delta}$  is closely related to the ordered pattern of Na ions which form one dimensional chains below room temperature \cite{Huang}. Because both of the one-dimensional Na order and the in-plane antiferromagnetic order are observed only for Na$_{0.5}$CoO$_2$, the effect of the Na order on the CoO$_2$ plane is considered to play crucial role for the antiferromagnetic order. The purpose of this paper is to investigate the electronic state of the CoO$_2$ plane, particularly focused on the effect of the Na order on the antiferromagnetic order at $x=0.5$. For this purpose, we take into account of an effective one-dimensional potential on the CoO$_2$ plane due to the effect of the Na order. 

\section{Model}
To investigate the electronic states of the CoO$_2$ plane in the layered cobalt oxides Na$_{x}$CoO$_2$, we employ the two dimensional triangular lattice $d$-$p$ model which includes 11 orbitals: 
 $d_{xy}$, $d_{yz}$, $d_{zx}$, $d_{x^2-y^2}$, $d_{3z^2-r^2}$ of Co
 and $p_{1x}$, $p_{1y}$, $p_{1z}$ ($p_{2x}$, $p_{2y}$, $p_{2z}$) of O
 in the upper (lower) side of a Co plane. 
The Hamiltonian is given by,
\begin{eqnarray}
\label{Model}
H
&=&
H_{p} + H_{dp} + H_{d} \,,
\\
H_{p}
&=&
\varepsilon_{p} \sum_{{\bi k},j,l,\sigma}
p_{{\bi k},j,l,\sigma}^{\dagger} p_{{\bi k},j,l,\sigma}
+ \sum_{{\bi k},j,j',l,l',\sigma} t_{{\bi k},j,j',l,l'}^{pp}
p_{{\bi k},j,l,\sigma}^{\dagger} p_{{\bi k},j',l',\sigma} \,,
\nonumber
\\
H_{dp}
&=&
\sum_{{\bi k},j,l,m} (t_{{\bi k},j,l,m}^{pd}
p_{{\bi k},j,l,\sigma}^{\dagger} d_{{\bi k},m,\sigma}
+ {\rm h.c.}) \,,
\nonumber
\\
H_{d}
&=&
\sum_{{\bi n} ,m,\sigma} \varepsilon_{{\bi n},m}^{d}
d_{{\bi n} ,m,\sigma}^{\dagger} d_{{\bi n} ,m,\sigma}
+ \sum_{{\bi k},m,m',\sigma} t_{{\bi k},m,m'}^{dd}
d_{{\bi k},m,\sigma}^{\dagger} d_{{\bi k},m',\sigma}
\nonumber
\\
&&
+ U \sum_{{\bi n} ,m}
d_{{\bi n} ,m,\uparrow}^{\dagger} d_{{\bi n} ,m,\uparrow}
d_{{\bi n} ,m,\downarrow}^{\dagger} d_{{\bi n} ,m,\downarrow}
\nonumber
\\
&&
+ U'\sum_{{\bi n} ,m \neq m'}
d_{{\bi n} ,m,\uparrow}^{\dagger} d_{{\bi n} ,m,\uparrow}
d_{{\bi n} ,m',\downarrow}^{\dagger} d_{{\bi n} ,m',\downarrow}
\nonumber
\\
&&
+ \left( U' - J \right) \sum_{{\bi n} ,m>m',\sigma}
d_{{\bi n} ,m,\sigma}^{\dagger} d_{{\bi n} ,m,\sigma}
d_{{\bi n} ,m,\sigma}^{\dagger} d_{{\bi n} ,m,\sigma}
\nonumber
\\
&&
+ J \sum_{{\bi n} ,m \neq m'}
d_{{\bi n} ,m,\uparrow}^{\dagger} d_{{\bi n} ,m',\uparrow}
d_{{\bi n} ,m',\downarrow}^{\dagger} d_{{\bi n} ,m,\downarrow}
\nonumber
\\
&&
+ J' \sum_{{\bi n} ,m \neq m'}
d_{{\bi n} ,m,\uparrow}^{\dagger} d_{{\bi n} ,m',\uparrow}
d_{{\bi n} ,m,\downarrow}^{\dagger} d_{{\bi n} ,m',\downarrow} \,,
\nonumber
\end{eqnarray}
where
 $d_{{\bi k},m,\sigma}^{\dagger}$ ($d_{{\bi n},m,\sigma}^{\dagger}$)
 is a creation operator for a cobalt $3d$ electron with wave vector ${\bi k}$ (site ${\bi n}=(n_x,n_y)$), orbital $m$ and spin $\sigma$, 
 and
 $p_{{\bi k},j,l,\sigma}^{\dagger}$ 
 is a creation operator for a oxygen $2p$ electron with  wave vector ${\bi k}$, site $j(=1,2)$, orbital $l$ and spin $\sigma$, 
 respectively. 
In eq. (\ref{Model}), the transfer integrals $t_{{\bi k},j,j',l,l'}^{pp}$, $t_{{\bi k},j,l,m}^{pd}$, $t_{{\bi k},m,m'}^{dd}$, which are written by the Slater-Koster parameters, together with the atomic energies $\varepsilon_{p}$, $\varepsilon_{m}^{d}$ are determined so as to fit the tight-binding energy bands to the LDA bands for Na$_{0.5}$CoO$_2$ \cite{Singh}. 
Effects of the Coulomb interaction at a Co site: 
 the intra- and inter-orbital direct terms $U$ and $U'$, 
 the exchange coupling $J$ and the pair-transfer $J'$ \cite{Mizokawa}. 
Here and hereafter, we assume the rotational invariance yielding the relations: $U'=U-2J$ and $J=J'$. 
We also consider the effect of the one-dimensional Na order, 
 in which Na ions form one dimensional chains,
 by taking into account of an effective one-dimensional potential
 on the CoO$_2$ plane: 
\begin{eqnarray}
\varepsilon_{{\bi n},m}^{d}
=
\left\{
\begin{array}{ll}
\varepsilon_{m}^{d} - \Delta \varepsilon_{d} 
& \qquad $for odd $n_y$ (on Na ordered line) $ \\
\varepsilon_{m}^{d} + \Delta \varepsilon_{d} 
& \qquad $for even $n_y$ (out of Na ordered line),$ \\
\end{array}
\right.
\end{eqnarray}
with the effective potential $\Delta \varepsilon_{d}$ due to the Na order. 
Hereafter, in order to consider the both Na order and the antiferromagnetic order, we use an extended unit cell including four Co atoms and eight O atoms, together with the magnetic Brillouin zone which is $\frac14$ of the original Brillouin zone.

\section{Result}
Now we discuss the antiferromagnetism in the case with $x=0.5$ within the Hartree-Fock approximation \cite{Mizokawa}. In this study, we assume that the order parameters are diagonal with respect to the orbital $m$ and the spin $\sigma$.

\begin{figure}[h]
\includegraphics[scale =0.4,angle=-90]{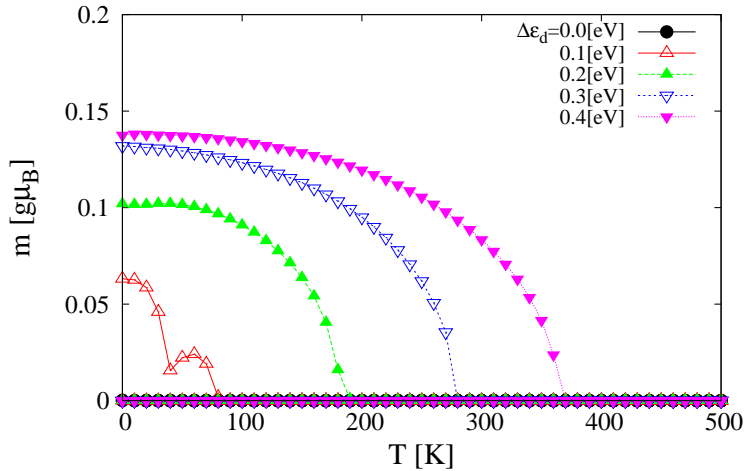}
\caption{
Sublattice magnetization $m$ as a function of the temperature $T$ for various values of the effective potential due to the Na order $\Delta \varepsilon_{d}$ at $U=1.5$ eV and $J=0.15$ eV. 
}
\label{fig-1}
\end{figure}

Figure \ref{fig-1} shows the sublattice magnetization $m$ as a function of the temperature $T$ for various values of the effective potential $\Delta \varepsilon_{d}$ at $U=1.5$ eV and $J=0.15$ eV. In the absence of the Na order $\Delta \varepsilon_{d} = 0$, no antiferromagnetic transition is observed for small values of $U$ and $J$ as the Fermi surface nesting with wave vector ${\bi q}=(\frac12,\frac12)$ corresponding to the in-plane antiferromagnetism is weak due to the strong frustration effect in the triangular lattice. With increasing $\Delta \varepsilon_{d}$, the antiferromagnetic transition temperature together with the sublattice magnetization increases. Due to the effect of the one-dimensional potential, the band structure becomes quasi one-dimensional and the Fermi surface nesting is enhanced. Then, the antiferromagnetic ordered state appears in the presence of the Na order. In this case, the sublattice magnetization exclusively comes from the Co sites on the Na ordered line. We note that the antiferromagnetic state is realized in the both cases with and without the small  Fermi surface pockets predicted  by the LDA calculation \cite{Singh} but not observed in the ARPES experiments \cite{Yang}. 

\begin{figure}[h]
\includegraphics[scale =0.6,angle=0]{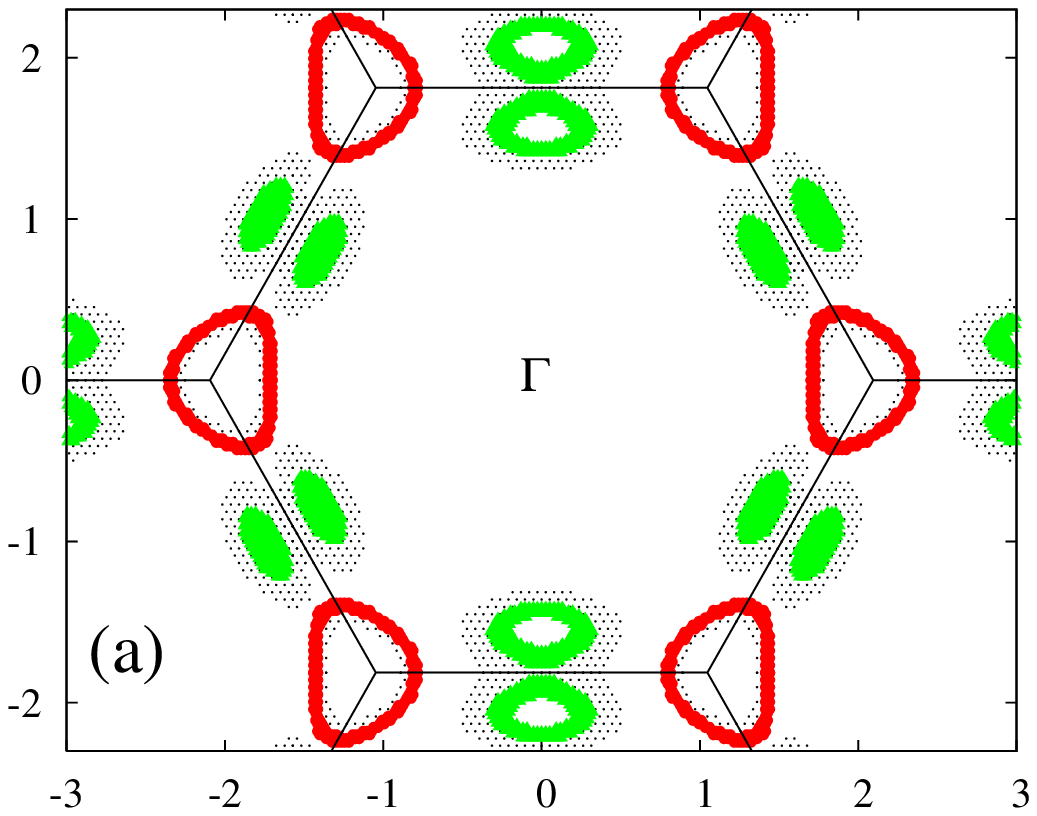}
\includegraphics[scale =0.65,angle=0]{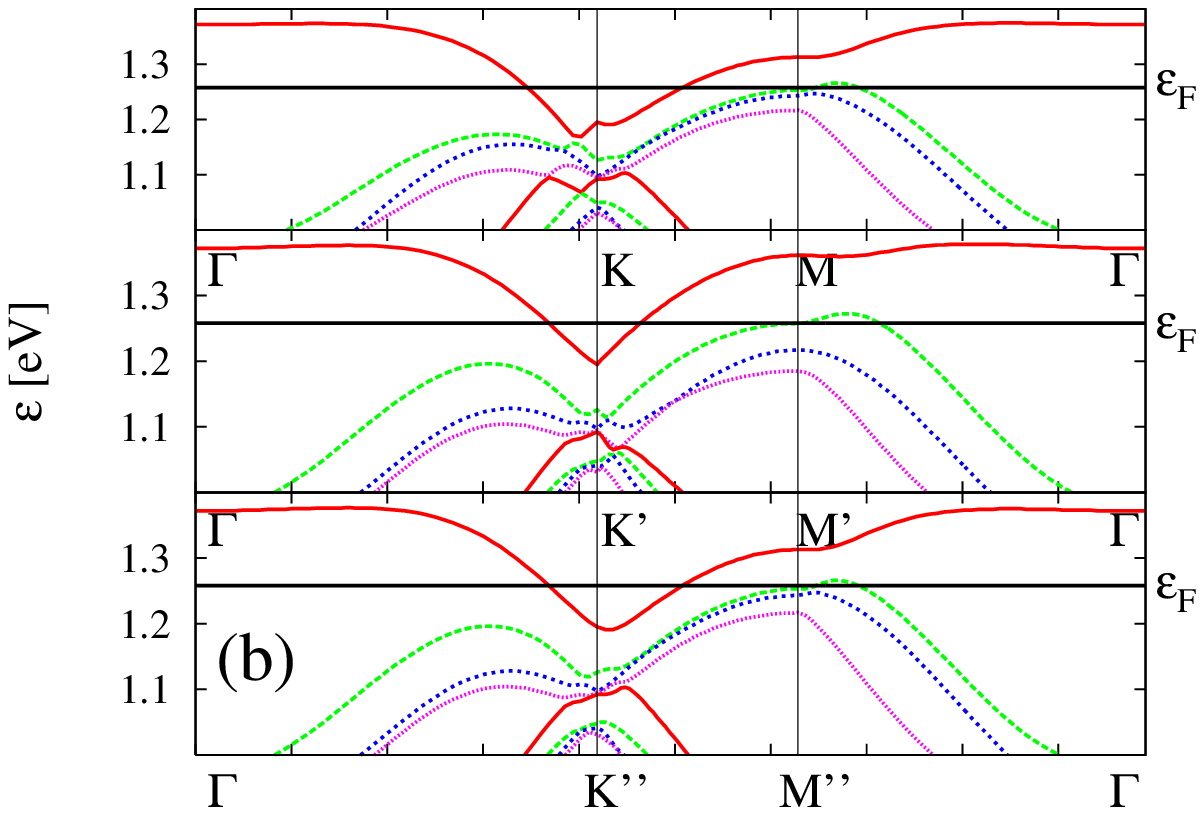}
\caption{
Fermi surfaces (a) and band structure (b) in the magnetic Brillouin zone for the metallic antiferromagnetic state at $U=1.5$ eV, $J=0.15$ eV, $\Delta \varepsilon_{d} = 0.2$ eV and $T=0$. 
}
\label{fig-2}
\end{figure}

\begin{figure}[h]
\includegraphics[scale =0.6,angle=0]{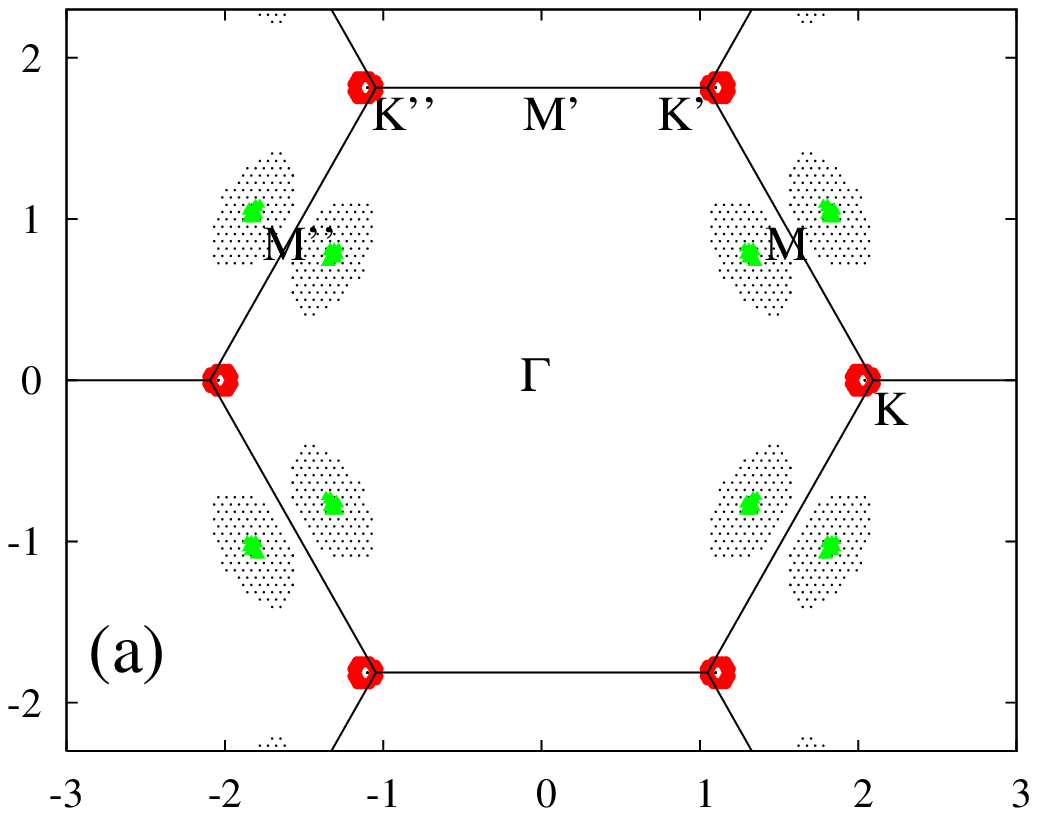}
\includegraphics[scale =0.65,angle=0]{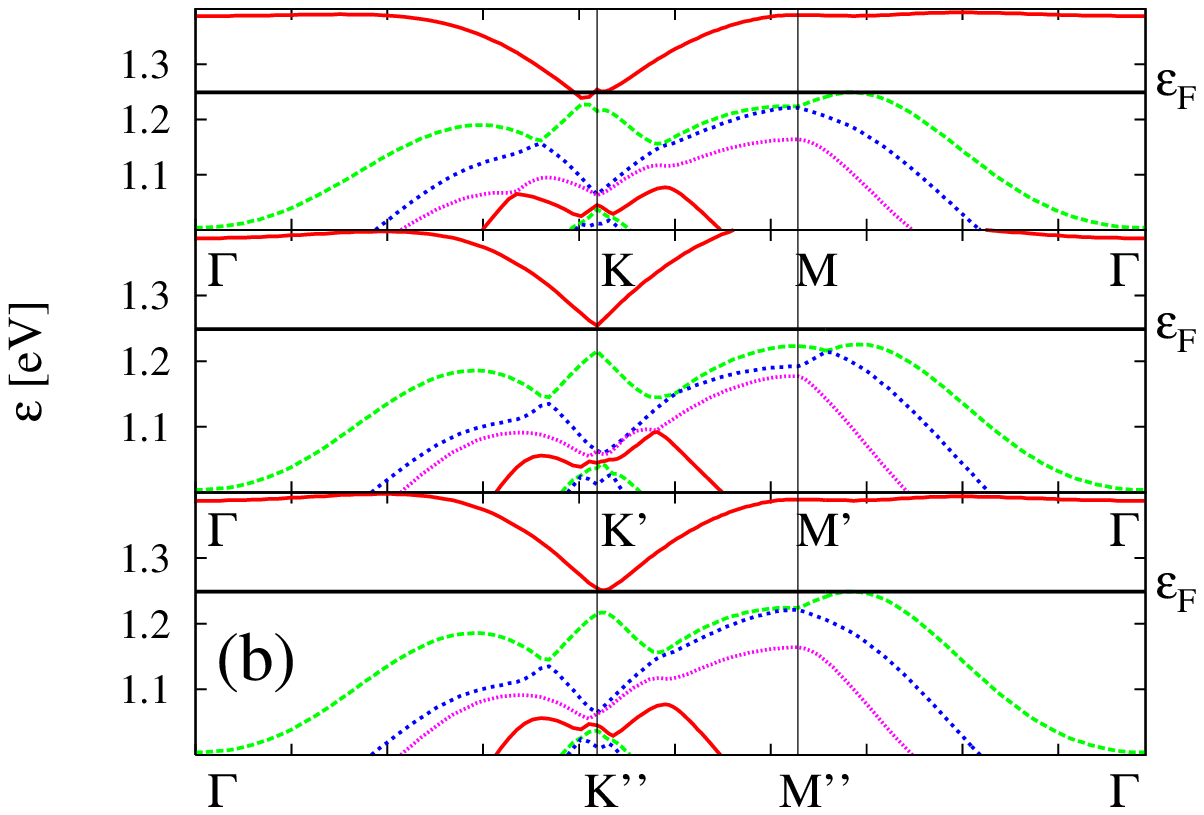}
\caption{
Fermi surfaces (a) and band structure (b) in the magnetic Brillouin zone for the semimetallic antiferromagnetic state at $U=2.0$ eV, $J=0.2$ eV, $\Delta \varepsilon_{d} = 0.2$ eV and $T=0$. 
}
\label{fig-3}
\end{figure}

\begin{figure}[h]
\includegraphics[scale =0.6,angle=0]{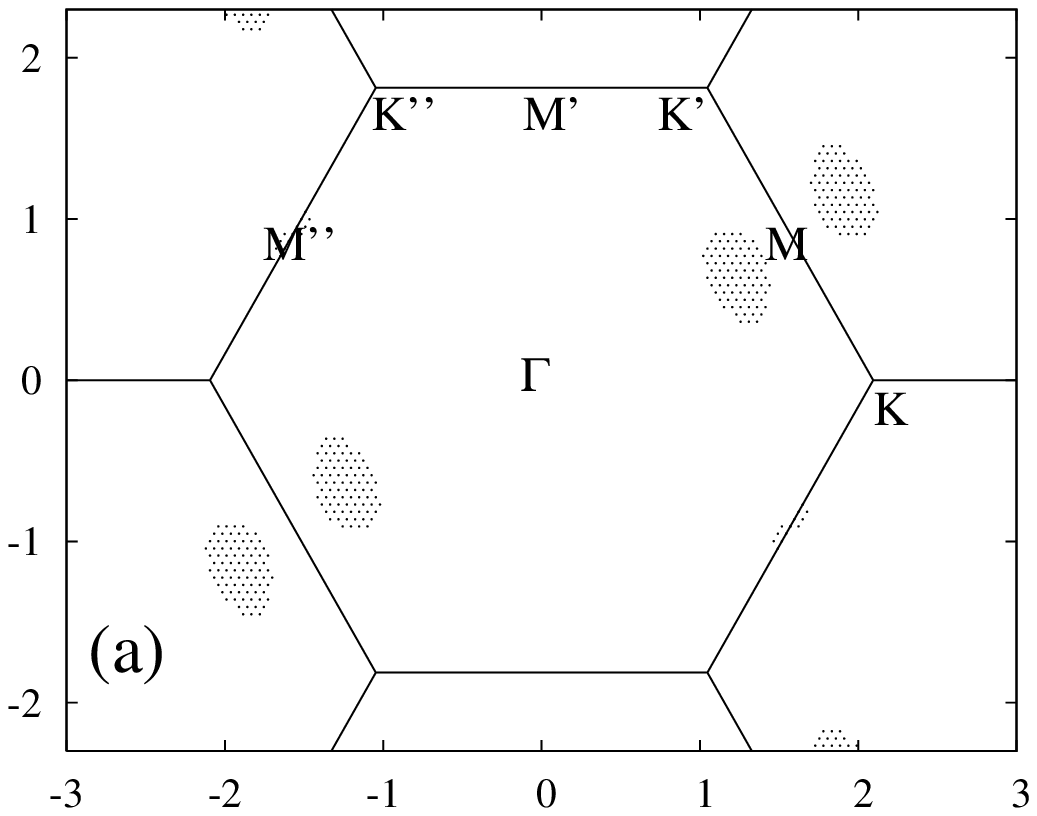}
\includegraphics[scale =0.65,angle=0]{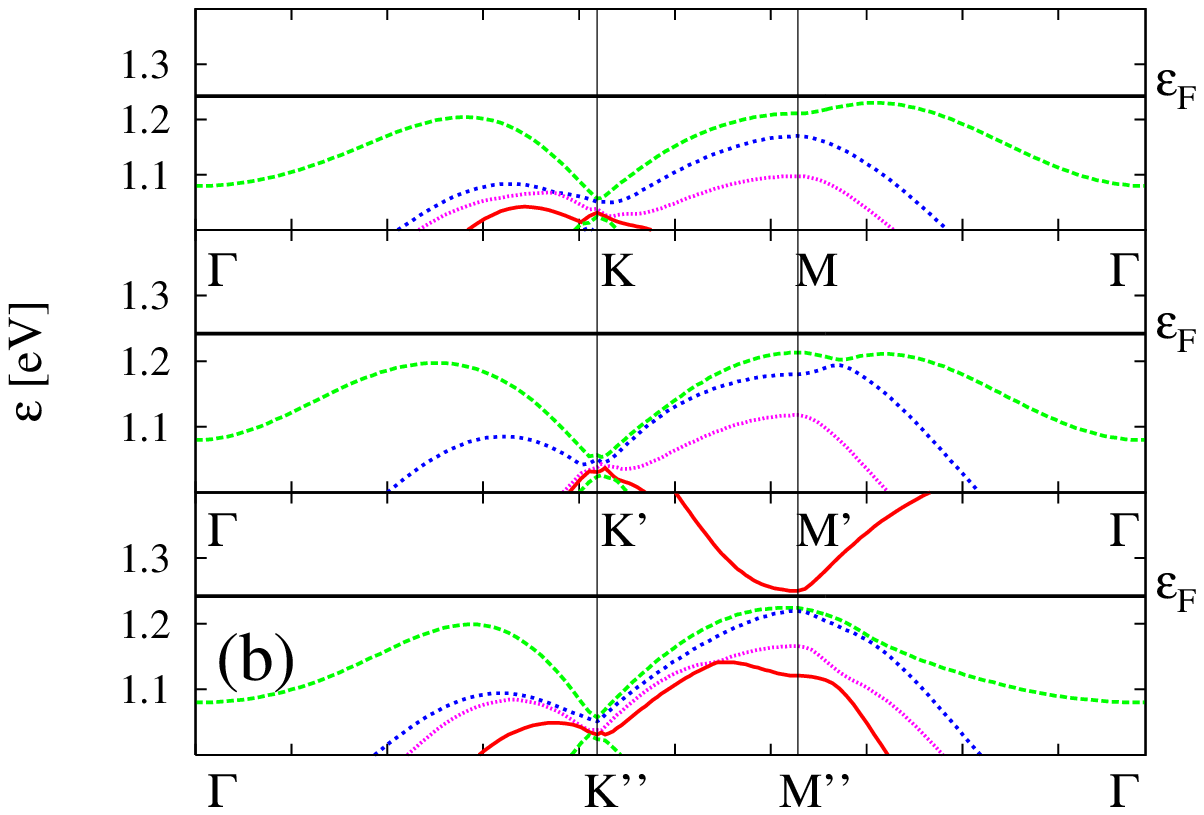}
\caption{
Fermi surfaces (a) and band structure (b) in the magnetic Brillouin zone for the insulating antiferromagnetic state at $U=2.0$ eV, $J=0.2$ eV, $\Delta \varepsilon_{d} = 0.4$ eV and $T=0$. 
}
\label{fig-4}
\end{figure}

In Figure \ref{fig-2}, we plot the Fermi surfaces and the band structure in the antiferromagnetic state at $T=0$ for $U=1.5$ eV, $J=0.15$ eV with $\Delta \varepsilon_{d} = 0.2$ eV. In this antiferromagnetic state, the system is metallic where the electron like Fermi surfaces around K points together with the hole like Fermi surfaces around M points yield a large value of the density of states $N_F$ at the Fermi level. On the other hand, when we increase the Coulomb interactions as $U=2.0$ eV and $J=0.2$ for the fixed value of $\Delta \varepsilon_{d} = 0.2$ eV, the ground state becomes the semimetallic antiferromagnetic state with tiny Fermi surface pockets around K points yielding a tiny $N_F$ as shown in Figure \ref{fig-3}. 
Furthermore, when  $U=2.0$ eV and $J=0.2$ with larger values of $\Delta \varepsilon_{d} \ge 0.35$ eV, the ground state becomes the insulating antiferromagnetic state with a finite energy gap as shown in Figure \ref{fig-4}. In this case, the sublattice magnetization from the Co sites out of the Na ordered line is larger than that from the Co sites on the Na ordered line, which is consistent with the magnetic structure proposed by Yokoi {\it et al.} \cite{Yokoi}.

\section{Summary}
We investigated the electronic state in the CoO$_2$ plane of the layered cobalt oxides Na$_{x}$CoO$_2$ by using the 11 band $d$-$p$ model on a two-dimensional triangular lattice within the Hartree-Fock approximation. 
What we found are: the effect of the one-dimensional potential $\Delta \varepsilon_{d}$ due to the Na order enhances the Fermi surface nesting resulting in the antiferromagnetic ordered state for $x=0.5$. There are three types of the antiferromagnetic states: (1) the metallic one for small values of $U$ and $\Delta \varepsilon_{d}$, (2) the semimetallic one for large $U$ with small $\Delta \varepsilon_{d}$ and (3) the insulating one for large values of $U$ and $\Delta \varepsilon_{d}$. Although the insulating antiferromagnetic state seems to be consistent with the experimentally observed ordered state below $T_{c2}$, we need further investigation especially on the mechanism of the metal-insulator transition at $T_{c2}$.


\section*{Acknowledgments}
The authors thank
 M. Sato, Y. Kobayashi, M. Yokoi and T. Moyoshi
 for many useful comments and discussions.
This work was performed under the interuniversity cooperative Research program of the Institute for Materials Research, Tohoku University, and was supported by the Grant-in-Aid for Scientific Research from the Ministry of Education, Culture,  Sports, Science  and Technology.

\end{document}